\documentclass[letterpaper, 10 pt, conference]{ieeeconf} 

\IEEEoverridecommandlockouts 

\usepackage{mathtools}
\usepackage{bbm}
\usepackage{lipsum}
\usepackage{comment}
\usepackage{amsfonts}
\usepackage{amsmath}

\usepackage{amssymb}
\usepackage{bm}
\usepackage[normalem]{ulem}
\usepackage{graphicx}
\usepackage{romannum}
\usepackage{graphics}
\usepackage{xcolor}
\usepackage{cite}
\usepackage{algorithm}
\usepackage{algpseudocode}
\usepackage{import}
\usepackage{balance}
\usepackage{algpseudocode}

\usepackage{amsmath,,amssymb,amsfonts,bm}
\usepackage[normalem]{ulem}
\usepackage{graphicx}
\usepackage{romannum}
\usepackage{amssymb}
\usepackage{graphics}
\let\labelindent\relax
\usepackage{enumitem}
\usepackage{xcolor}
\usepackage{cite}
\usepackage{algorithm}
\usepackage{algpseudocode}

\newcommand{\vi}[1]{\textcolor{violet}{[Vi: #1]}}

\newcommand{\edit}[1]{\textcolor{blue}{#1}}

\algrenewcommand{\algorithmiccomment}[1]{%
\hfill\parbox[t]{0.45\linewidth}{\raggedright\fontsize{8}{8}\selectfont\texttt{\%}{ #1}}}

\newtheorem{definition}{Definition}

\newtheorem{rem}{Remark}
\newtheorem{prop}{Proposition}

\usepackage{import}
\newcommand{\Attacker}{Attacker}

\newcommand{\Defender}{Defender}
\newcommand{\graph}{\mathcal{G}}
\newcommand{\nodeset}{\mathcal{V}}
\newcommand{\edgeset}{\mathcal{E}}
\newcommand{\weight}{w}
\newcommand{\Aactsset}{\mathcal{A}^{\theta}}
\newcommand{\Dact}{a^D}

\newcommand{\Defs}{\pi}
\newcommand{\Atts}{\gamma}

\newcommand{\Defsset}{\Pi}
\newcommand{\Attsset}{\Gamma}

\newcommand{\restrictedH}{\hat{\mathcal{H}}}

\newenvironment{Proof}{\par\noindent\textit{Proof.}\ }{\hfill$\square$\par}

\usepackage{balance} 

\title{Linear Programming Approach to\\ Deceptive Path Planning Game with Goal Selection}

\author{Violetta Rostobaya$^{1}$, Yue Guan$^{2}$, James Berneburg$^{1}$ and Daigo Shishika$^{1}$
\thanks{
We gratefully acknowledge the support of  ARL grant ARL DCIST CRA
W911NF1720181 and ARL grant W911NF2520011.}
\thanks{$^1$Violetta Rostobaya, James Berneburg and Daigo Shishika are with College of Engineering and Computing at George Mason University, Emails: {\tt\footnotesize $\{$vrostoba,jbernebu,dshishik$\}$@gmu.edu}}
\thanks{$^2$Yue Guan is with the School of Aerospace Engineering, Georgia Institute of Technology, Email: {\tt\footnotesize yguan44@gatech.edu}}
}

\begin{document}

\makeatletter
\makeatother

\maketitle
\thispagestyle{empty}
\pagestyle{empty}

\begin{abstract}
In adversarial settings, a mobile agent may strategically plan its motion to influence an opponent’s inference about its intended goal. 
We study deceptive path planning in a scenario where a mobile agent aims to reach a privately selected goal while an adversarial observer allocates limited defensive resources based on the observed trajectory. 
Unlike classical path-planning and goal-recognition approaches that model observers as passive inference process, our game-theoretic formulation models them as strategic decision-makers.
For the resulting dynamic asymmetric-information game, we develop an efficient solution method that combines a linear programming formulation with the Double Oracle algorithm. 
To evaluate performance, we introduce metrics that quantify both the risk and the effectiveness of deception and provide illustrative numerical examples. 

\end{abstract}
\section{Introduction}

\begin{figure}[t!]
\centering
\includegraphics[width = 0.48\textwidth]{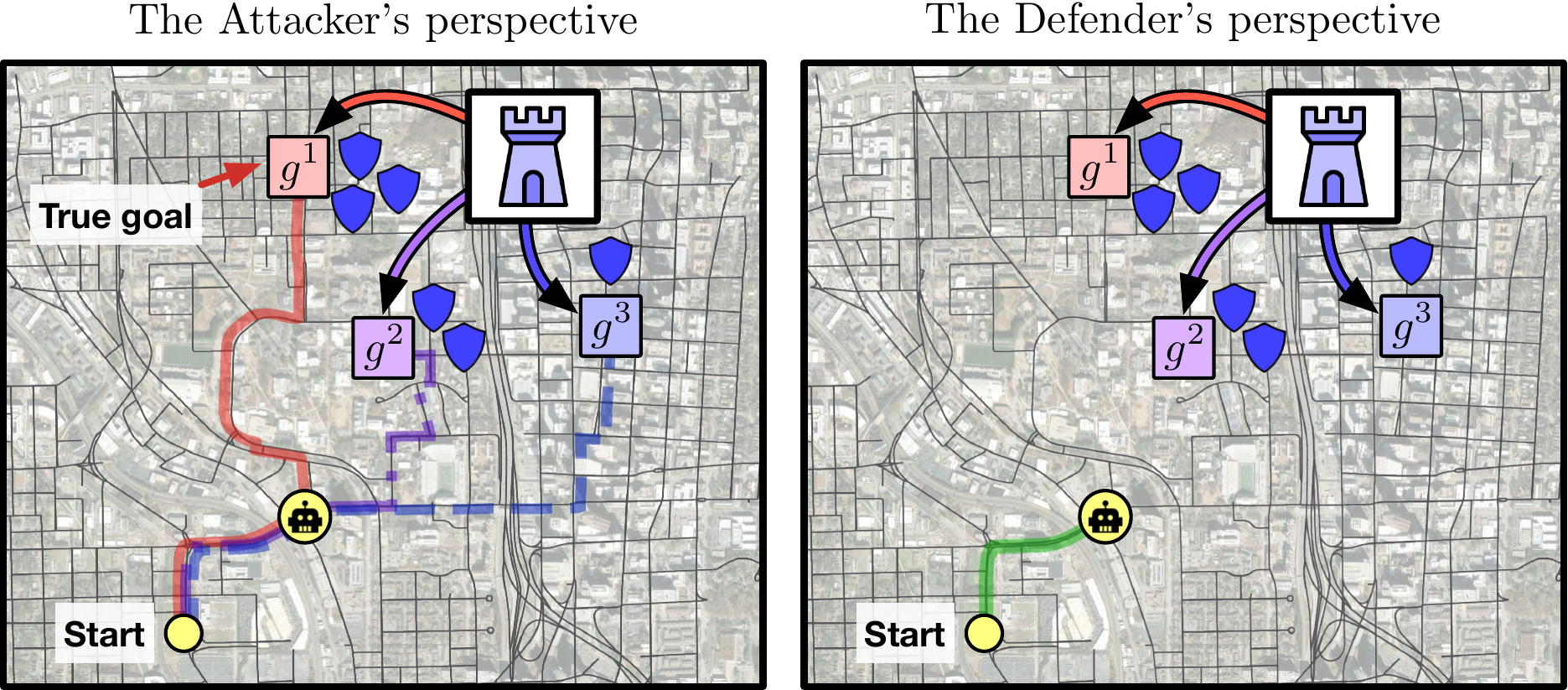}
\caption{The Attacker aims to reach its selected goal $g^1$. The Defender does not know the true goal, but it aims to infer the Attacker's intentions and allocate maximum amount of defensive resources to the true goal. }\label{fig: figure_1}
\end{figure}

\noindent
Deception plays an important role in domains such as military operations~\cite{lloyd2003art}, security planning~\cite{almeshekah2016cyber}, and autonomous systems operating in contested environments~\cite{nikitas2022deceitful}. 
In this paper, we study deception in the context of motion planning, where a mobile agent seeks to reach a privately selected target while a surveilling opponent allocates limited defensive resources to candidate goals as illustrated in Fig.~\ref{fig: figure_1}.

This setting is closely related to \emph{goal recognition}~\cite{Wayllace2016GoalRD, Pereira2017LandmarkBasedHF,ramirez2011goal},
in which an observer infers a mobile agent’s intent from its trajectory, typically assuming that the agent follows a cost-minimizing plan. 
In \emph{deceptive path planning} (DPP), goal-recognition models are often used to specify how an observer infers an agent’s intent, against which the mobile agent optimizes its trajectory to reach its true goal while reducing inference accuracy~\cite{dragan2015deceptive,masters2017deceptive,Ornik2018DeceptionIO,karabag2021deception}.
However, these works often treat the steering of observer's beliefs as the primary objective without considering its effect on the observer's decision making.
We address this gap by formulating these deceiver-observer interactions as games with explicit models of observer's action.

Prior works on game-theoretic formulation of DPP considered different environments, information structures, and/or equilibrium concepts.
The initial work that introduced a game with a resource-allocating observer considered a grid-world environment with worst-case payoff~\cite{rostobaya2023deception}.
This work was later extended to general graph environments with zero-sum Bayesian formulation~\cite{rostobaya2025deceptive}, where the goal was randomly assigned to the agent with a known prior probability.
Both works assumed that the deceiver must reach one of two candidate goals (i.e., there are two possible \emph{types} of agents).
In contrast, we generalize the DPP game to a multi-goal scenario. 
Moreover, we allow the deceiver to select its goal at the start of the game, thereby eliminating the need to assume a prior probability distribution over goals.

Computational complexity is another important consideration for the DPP game.
Prior works have leveraged specific geometric structure to induce tractable solutions~\cite{rostobaya2023deception,rostobaya2025deceptive}.
More generally, dynamic imperfect-information games such as ours that allow game-tree representation may be solved using a linear programming (LP)~\cite{koller1996efficient} or bilinear program~\cite{guan2025strategic} under certain structural assumptions. 
Specifically, 
zero-sum asymmetric-information games can be solved with LP when the state is controlled solely by the informed player~\cite{JeffShammaLP_ZS_2014}. 
However, representing the game in extensive form leads to a game tree whose size grows exponentially with the planning horizon, making direct solution methods computationally challenging for large environments.

To address this computational challenge, we adapt the Double Oracle algorithm developed for \emph{finite} extensive-form games (XDO)~\cite{mcaleer2021xdo}, which incrementally expands the game tree until it is sufficient for an $\varepsilon$-Nash equilibrium. In particular, we combine this approach with the LP formulation~\cite{JeffShammaLP_ZS_2014}, yielding an iterative method that is well-suited for our multi-goal DPP game.
Leveraging the one-sided incomplete-information structure and a particular choice of \emph{default strategies}~\cite{Boansk2014AnED,mcaleer2021xdo}, we are able to solve DPP game with indefinite horizon (i.e., the termination time is not specified apriori) and compute an exact Nash equilibrium rather than an $\varepsilon$-Nash equilibrium.



The main contributions of this work are:
(i) formulation of a new variant of the DPP game with goal selection, (ii) solution of the game via an algorithm that integrates the LP formulation with the Double Oracle algorithm for extensive-form games; and
(iii) metrics to evaluate risk and effectiveness of emergent deceptive behavior, benchmarked against the complete-information setting.

\section{Preliminaries}
\noindent
We consider an asymmetric information deceptive path planning game between a mobile agent, \Attacker{}, and an observing agent, \Defender{}, as illustrated in Fig.~\ref{fig: figure_1}. 
The \Attacker{} moves in a graph environment containing multiple candidate goals. 
At the start of the game, the \Attacker{} selects a goal that it must reach.
The \Defender{}, lacking knowledge of the selected goal, must infer it based solely on observations of the \Attacker{}'s trajectory. 
The \Defender{} seeks to allocate defense resources to the selected goal before the \Attacker{} arrives. 
This setting naturally incentivizes the \Attacker{} to move in a way that conceals its true goal and misleads the \Defender{} into allocating resources to the wrong goals.

\paragraph*{Environment}
The graph is denoted as $\graph = \langle \nodeset, \edgeset, \weight \rangle$,
where $\nodeset$ is a finite set of nodes, and $\edgeset \subseteq \nodeset \times \nodeset$ is the set of edges. We use $\mathcal{N}(s)$ to represent the out-neighbors of $s$ on $\graph$.
The positive edge weight is given by the function $\weight: \edgeset \rightarrow \mathbb{R}_{>0}$.
The weight $w(s,s')$ represents the amount of defense resource the \Defender{} can allocate to one of the goals while the \Attacker{} is traversing through edge $(s,s')$.
The set of candidate goals is denoted by $\Theta \subseteq \nodeset$.
At the beginning of the game, the \Attacker{} selects a goal $\theta\in \Theta$.
An \Attacker{} that selects goal $\theta$ is said to be of type $\theta$.
Once the goal is selected, the \Attacker{} must commit to it for the remainder of the game; that is, the \Attacker{}'s type remains fixed.

\paragraph*{Dynamics}
The game state at time $t$ is given by the \Attacker{}'s position $s_t \in \nodeset$.
At $t=-1$, the \Attacker{} makes a pre-game decision and selects its type from the set $\Theta$. This choice is not revealed to the Defender and remains private to the Attacker till the end of the game. At $t\geq0$, the \Attacker{} with type $\theta$ chooses its next position 
$s_{t+1}$ from its action set:
\begin{equation*}
 \mathcal{A}^{\theta}(s_t)\coloneqq\begin{cases}
 \mathcal{N}(s_t), \; &\text{if}\; s_t\neq \theta,\\
 \{\theta\}, \; &\text{otherwise},
 \end{cases}
\end{equation*}
 i.e., if the Attacker reaches its selected goal, it stays there.

At each time step $t\geq 0$, the \Defender{} selects a single goal $\Dact_t \in \Theta$ to allocate resources to.
Choosing action $\Dact_t=i$ corresponds to allocating $\weight(s_t,s_{t+1})$ units of defense resource to goal $i\in \Theta$, where $(s_t,s_{t+1})$ is the edge traversed by the \Attacker{} at $t$.
Once the resource is allocated to a goal, it cannot be reallocated to other goals later on.


\paragraph*{Terminal condition} The game ends when the Attacker reaches its selected goal $\theta$. We denote $T^{\theta}$ to be the terminal time for the Attacker type $\theta$. 
Note that $T^{\theta}$ is a random variable conditioned on the Attacker $\theta$'s strategy:
\begin{equation*}
T^{\theta}=\min_{t\in \mathbb{Z}_{\geq 0}}\{t|s_t=\theta \}.
\end{equation*}

\paragraph*{Information structure}
The 
graph $\graph$ and the candidate goal set $\Theta$ are common information and are omitted in the information sets of both players.

We denote public history of the Attacker's positions as $h_t=(s_0,s_1,\dots,s_t)$ with the initial history $h_0=(s_0)$, and the set of all possible state histories at time $t$ as $\mathcal{H}_t$. 
We say that $h_m$ is a \emph{prefix} of a history $h_t$, denoted as $h_{m}\sqsubseteq h_{t}$, if $\exists\, h'$ such that $h_t = (h_m, h')$. That is, the first $m$ actions of $h_t$ coincide exactly with all the actions in $h_m$.

We denote the Attacker's information set at time $t=-1$ as $I^{A}_{t=-1}=(s_0)$, which informs the Attacker about its initial position. 
For $t\geq 0$, we consider simultaneous actions. 
The Attacker's information set is a tuple $I^A_t=(\theta,h_t)$, 
which informs the Attacker about its selected type $\theta$ and its current state history $h_t$. 
The set of possible Attacker's information sets is denoted as $\mathcal{I}^{A}_t$, where $\mathcal{I}^{A}_{-1}=\mathcal{V}$ and $\mathcal{I}^{A}_{t\geq 0}= \Theta \times \mathcal{H}_t$.
Since the Defender does not have access to $\theta$, its information set for $t\geq 0$ is simply $I^{D}_t=(h_t) \in \mathcal{I}^{D}_{t}=\mathcal{H}_t$. 

\paragraph*{Strategy sets}
We let $\Attsset$ be the set of admissible behavioral strategies of the \Attacker{}, and $\Atts=(\gamma_{\text{type}},(\gamma^{\theta})_{\theta\in\Theta})\in\Attsset$ to be a specific \Attacker{}'s strategy.
The \Attacker{}'s selection of a goal at $t=-1$ is given by a mapping $\Atts_{\text{type}}: \mathcal{I}^A_{-1} \rightarrow \Delta(\Theta)$, where $\Delta(\Theta)$ denotes an $|\Theta|$-dimensional simplex.
Policy $\Atts_{\text{type}}(\theta|s_0)$ determines the probability that the \Attacker{} selects goal $\theta$ given initial position $s_0$. 
For the rest of the game, i.e., for $t\geq 0$, the \Attacker{} type $\theta$ plays strategy $\gamma^\theta$, where $\gamma^\theta_{h_t}:\mathcal{I}^A_t\rightarrow\Delta(\mathcal{A}^{\theta}(\cdot))$.
The behavioral strategy $\gamma^\theta_{h_t}$ is a probability distribution over $\mathcal{A}^{\theta}(s_t)$. Thus, $\gamma^\theta_{h_t}(s')$ denotes the probability that the Attacker of type $\theta$ selects next position $s'\in \mathcal{A}^{\theta}(s_t)$ given history $h_t$.
Note that we restrict $\Gamma$ to be such that the Attacker type $\theta$ reaches its goal in finite time, i.e., $\mathbb{P}^{\gamma}(T^{\theta}<\infty)=1, \forall\theta \in \Theta$, and $\forall \gamma \in \Gamma$.

Similarly, we use $\Defsset$ to denote the set of \Defender{}'s admissible behavioral strategies, and we let $\Defs\in\Defsset$ to be its element, which provides the \Defender{}'s strategy $\Defs_{h_t}: \mathcal{I}^{D}_t \rightarrow \Delta(\Theta)$.
The strategy $\Defs_{h_t}(\Dact_t)$ denotes the probability that the \Defender{} allocates to goal $\Dact_t \in \Theta$ after observing history $h_t$.

\paragraph*{Probabilities}
We denote by $p^\theta_{h_t}\triangleq\mathbb{P}^{\gamma}(\theta|h_t)$ the probability that the Attacker is type $\theta$ given observed history~$h_t$. This probability is conditioned on the Attacker's strategy~$\gamma$.
Since the Attacker chooses its type at $t=-1$,
the elements of the initial probability $\textbf{p}_{h_0}\in \Delta(\Theta)$ is given by $ p^{\theta}_{h_0}=\gamma_{\text{type}}(\theta|s_0)$. 
For $t\geq0$ the probability $\textbf{p}_{h_t}$ is propagated via Bayes rule:
\begin{equation}\label{eq. belief bayes rule}
p^{\theta}_{h_{t+1}}=\frac{p^{\theta}_{h_t}\Atts^{\theta}_{h_t}(s_{t+1})}{\underset{\psi\in\Theta}{\sum} p^{\psi}_{h_t}\Atts^{\psi}_{h_t}(s_{t+1})},
\end{equation}
where $h_{t+1}=(h_t,s_{t+1})$.
\paragraph*{Stage cost}
The goals have initial allocation described by the vector $\mathbf{r}\in \mathbb{R}^{|\Theta|}_{\geq 0}$. 
We let $r^\theta\coloneqq[\mathbf{r}]_\theta$ denote the component corresponding to goal~$\theta$. 
The initial allocation can be viewed as the preparedness level assigned to each goal for countering a potential attack.
The \Attacker{}'s stage cost at $t$ depends on the traversed edge $(s_t,s_{t+1})$, the \Attacker{}'s type $\theta$ and the \Defender{}'s action $a^D_t$:
\begin{equation}
 l^{\theta}(s_t,s_{t+1},\Dact_t)=\begin{cases}
w(s_{t},s_{t+1}),&\text{if } \Dact_t=\theta, 
\\
0,&\text{otherwise}.
 \end{cases}
\end{equation}
 In other words, the \Attacker{} $\theta$ is penalized only when the \Defender{} allocates to the true goal $\Dact_t=\theta$. 

\paragraph*{Objective function}
We consider a zero-sum game where the \Attacker{} aims to minimize the expected cumulative cost, while the \Defender{} aims to maximize it. 
Once the Attacker chooses its goal $\theta$, we can express the Attacker $\theta$'s total expected cost as follows:
\begin{equation*}
J^\theta(\pi,\gamma^{\theta};s_0,\mathbf{r})=\mathbb{E}^{\gamma^{\theta},\pi} \left[ \sum_{t=0}^{T^{\theta}-1} l^\theta(s_t,s_{t+1},\Dact_t) \right]+r^{\theta}.
\end{equation*}
Then the overall expected cost of the \Attacker{} is:
\begin{align}\label{eq. total payoff equation}
J(\pi,\gamma;s_0,\mathbf{r})&=\mathbb{E}^{\gamma,\pi} \left[J^\theta(\pi,\gamma^{\theta};s_0,\mathbf{r}) \right]\notag\\
 &=\sum_{\theta\in \Theta} \gamma_{\text{type}}(\theta|s_0)J^\theta(\pi,\gamma^{\theta};s_0,\mathbf{r}). 
\end{align}
%
Note that the Attacker can employ a type-specific strategy for each realized type, whereas the Defender must deploy the same strategy across all possible Attacker types.

We consider an equilibrium concept defined as follows:
\begin{definition}
A strategy profile $(\Atts^*,\Defs^*)$ constitutes a Nash equilibrium (NE) if for all $\Atts\in \Attsset$ and $\Defs\in \Defsset$ it satisfies:
\begin{equation}\label{eq. Nash}
J(\Defs,\Atts^*;s_0,\mathbf{r}) \leq J(\Defs^*,\Atts^*;s_0,\mathbf{r})\leq J(\Defs^*,\Atts;s_0,\mathbf{r}).
\end{equation}
\end{definition}
\noindent
The next section describes the proposed solution method.

\section{Solution Method}
\noindent
The DPP game presented above is a zero-sum stochastic game with asymmetric information in which the state is controlled only by the informed agent, the Attacker. This setting allows the use of the LP method presented in~\cite{JeffShammaLP_ZS_2014}.
However, a key difference is that while \cite{JeffShammaLP_ZS_2014} considers games with known and fixed terminal time, our DPP game has free terminal time: i.e., the game termination depends on the Attacker's strategy.
Therefore, in Section~\ref{subsection restricted game}, we firstly provide an LP method to solve a \emph{restricted game}, which is a prefix-closed truncation of the original extensive-form game to a finite set of histories.
In Section~\ref{subsection unrestricted game}, we secondly provide a Double-Oracle algorithm inspired by~\cite{Boansk2014AnED,mcaleer2021xdo} that iteratively expands the restricted game to find the solution for the overall unrestricted game.

\subsection{Solution for a restricted game.}\label{subsection restricted game} 
\noindent
We first analyze a \emph{restricted} game, which is a prefix-closed truncation of the original extensive-form game to a set of histories $\restrictedH\subset \bigcup^{\infty}_{t=0}\mathcal{H}_t$. By definition of prefix-closedness, if a history $h_{t+1}$ belongs to $ \restrictedH$, then its prefix $h_t$ also belongs to $ \restrictedH$. 
We obtain the initial restricted game by propagating the game tree up to a fixed time horizon $\tau > 0$.
Accordingly, we define the initial set of histories $\restrictedH$ as the set of histories reachable by time $\tau$, i.e., $\restrictedH \coloneqq \bigcup_{t=0}^{\tau} \mathcal{H}_t$.

We let 
$\overline{\mathcal{H}}^\theta\coloneqq\{h_t\in \restrictedH| s_m\neq \theta, \; \forall h_m \sqsubseteq h_t\}$
to be a set of non-terminal histories for Attacker $\theta$ in the unrestricted game. 
Let $\mathcal{H}_f^\theta \subset \restrictedH$ denote the set of \emph{final histories} for Attacker $\theta$ in the restricted game, defined as the histories that either (i) reach the terminal condition (i.e., the corresponding goal) in the original game, or (ii) those that lie on the truncation frontier $t=\tau$.
 Based on the above two cases, for each type $\theta$, we partition $\mathcal{H}_f^\theta$ into the subset that ends the restricted game at the selected goal $\theta$, i.e.,
$\mathcal{H}^{\theta*}_f\coloneqq \restrictedH \setminus\overline{\mathcal{H}}^\theta $,
and its complement that ends at any other state, i.e., $
\overline{{\mathcal{H}}}_f^\theta \coloneqq \mathcal{H}_f^{\theta} \setminus \mathcal{H}^{\theta*}_f$.
For the latter, the original game continues beyond $\tau$. 
Any history $h_t\in \overline{\mathcal{H}}_f^\theta$ is treated as terminal with an exogenously specified continuation cost for Attacker $\theta$ denoted as $\mathcal{U}^{\theta}_{h_t}$.
The method for computing this continuation cost is described in Section~\ref{subsection unrestricted game}.

For all non-terminal histories $h_t\in \overline{\mathcal{H}}^\theta$, we can represent the stage game as a matrix $L^{\theta}_{h_t} \in \mathbb{R}^{|\Theta| \times |\Aactsset(s_t)|}$: 
\begin{align}\label{eq. stage cost matrix}
\left[L^{\theta}_{h_t}\right]_{a^D_t,s_{t+1}}
=
\begin{cases}
l^{\theta}(s_t,s_{t+1},a^D_t)+\mathcal{U}^{\theta}_{h_{t+1}}, \text{if } h_t \in \overline{\mathcal{H}}_f^\theta,\\
l^{\theta}(s_t,s_{t+1},a^D_t)+r^{\theta}, \text{if } h_t = h_0,\\
l^{\theta}(s_t,s_{t+1},a^D_t), \text{otherwise,}
\end{cases}
\end{align}
where $a^D_t\in \Theta$ is the row player's (Defender's) action, and $s_{t+1}\in \mathcal{A}^{\theta}(s_t)$ is the column player's (Attacker's) action.
For any terminal history that ends the overall unrestricted game for Attacker $\theta$, i.e., $h_t\in \mathcal{H}^{\theta*}_f$, we have $L^{\theta}_{h_t}=\mathbf{0}^{|\Theta|\times |\Aactsset(s_t)|}$.

We denote the information sets of the Attacker (resp. the Defender) in the restricted game for $t=-1$ as $\hat{\mathcal{I}}^{A}_{-1}=\mathcal{V}$
and for $t\geq0$ as $\hat{ \mathcal{I}}^{A}_{t}=\mathcal{H}_t \times \Theta$, (resp. for $t\geq 0$ as $\hat{ \mathcal{I}}^{D}_{t}=\mathcal{H}_t$), where $\mathcal{H}_t \in \restrictedH{}$.
We let $\hat \Gamma$ be the Attacker's admissible strategy set in the restricted game, and its element be an Attacker's strategy $\hat \gamma\coloneqq(\hat \gamma_{\text{type}},(\hat \gamma^{\theta})_{\theta \in \Theta})$. 
Here $\hat \gamma_{\text{type}}:\hat{\mathcal{I}}^{A}_{-1}\to \Delta{(\Theta)}$, and $\hat \gamma^{\theta}_{h_t}:\hat{\mathcal{I}}^{A}_{t}\to \Delta{(\Aactsset(\cdot))}$.
Similarly we define the Defender's strategy set in the restricted game as $\hat \Pi$ and its element $\hat \pi$, where $\hat \pi_{h_t}: \hat{\mathcal{I}}^{D}_{t}\to \Delta{(\Theta)}$.

The quantity $\mathcal{J}_{h_t}^{*}(\mathbf{p}_{h_t},\mathbf{r})$ represents the equilibrium expected cost from history $h_t$ onward.
We define the equilibrium expected continuation cost starting from history $h_t$ given the probability $\mathbf{p}_{h_t}$ over the Attacker types as the following min-max problem:
\begin{equation}\label{eq. equilibrium expected cost}
 \mathcal{J}_{h_t}^{*}(\mathbf{p}_{h_t},\mathbf{r})=\max_{\hat \pi} \min_{\hat\gamma} \sum_{\theta\in \Theta}p^{\theta}_{h_t}\mathcal{J}^{\theta}_{h_t}(\hat \pi, \hat \gamma^{\theta},\mathbf{r}),
\end{equation}
where the probability $\mathbf{p}_{h_t}$ is defined by \eqref{eq. belief bayes rule}, $\mathbf{p}_{h_0}=\gamma_{\text{type}}$. 
In~\eqref{eq. equilibrium expected cost} term $\mathcal{J}^{\theta}_{h_t}(\hat \pi,\hat \gamma^{\theta},\mathbf{r})$ is the expected cost for the Attacker type $\theta$ at history $h_t$, when the strategies in the restricted game are $\hat \pi$ and $\hat \gamma^{\theta}$:
\begin{subequations}
\begin{align}
 & \mathcal{J}^{\theta}_{h_t}(\hat \pi,\hat \gamma^{\theta},\mathbf{r})=\hat \pi_{h_t}^{\top}L^\theta_{h_t}\hat \gamma^{\theta}_{h_t} +\\
&\sum_{s_{t+1}\in \mathcal{A}^{\theta}(s_t)}\hat \gamma^{\theta}_{h_t}(s_{t+1})\mathcal{J}_{h_{t+1}}^{\theta}(\hat \pi,\hat\gamma^{\theta},\mathbf{r}), \; \forall h_{t} \in{\restrictedH}\setminus {\mathcal{H}}^\theta_f , \notag\\
&\mathcal{J}^{\theta}_{h_t}(\hat \pi,\hat \gamma^{\theta},\mathbf{r})=\hat \pi_{h_t}^{\top}L^\theta_{h_t}\hat \gamma^{\theta}_{h_t}, \forall h_t \in {\mathcal{H}}^\theta_f.
\end{align}
\end{subequations}

By directly applying the results of \cite{JeffShammaLP_ZS_2014}, the Attacker's strategy for a restricted game can be found via the following primal LP:
\begin{gather}\label{eq. Attacker's LP}
\min_{\substack{\hat\gamma_{\text{type}}, \; (z^{\theta}_{h_t})_{\theta \in \Theta},\\v_{h_{t}},\forall h_t\in\restrictedH}} \quad\sum_{h_t\in\restrictedH }v_{h_t}\\
 \text{s.t. } \mathbf{1}^{\top}z^{\theta}_{h_0}=\hat \gamma_{\text{type}}(\theta|s_0),\notag\\
 \mathbf{1}^{\top} \hat\gamma_{\text{type}}=1,\;\hat \gamma_{\text{type}}\geq \mathbf{0},\notag\\
 \text{and }\forall h_{t+1}=(h_{t},s_{t+1}), \; h_t, h_{t+1}\in \restrictedH:\notag\\
\sum_{\theta\in{\Theta}} L^{\theta}_{h_{t}}z^{\theta}_{h_{t}}\leq v_{h_{t}}\mathbf{1},\notag\\
 \mathbf{1}^{\top}z^\theta_{h_{t+1}}= z^\theta_{h_{t}}(s_{t+1}),\;
 z^{\theta}_{h_{t}}\geq \mathbf{0} \; \notag
\end{gather}
Here $z^{\theta}_{h_t}=p^{\theta}_{h_t}\hat \gamma^{\theta}_{h_t}$.
The solution for the Attacker's strategy in the restricted game is given by the optimal minimizing arguments $\hat \gamma^{*}_{\text{type}}$ and $z^{\theta*}_{h_t}$ of~\eqref{eq. Attacker's LP}. Note that $\hat \gamma^{*}=(\hat\gamma^{*}_{\text{type}},(\hat \gamma^{\theta*})_{\theta \in \Theta})$, where $\hat \gamma^{\theta*}_{h_t}$ for $h_t \in \restrictedH$ is given by:
\begin{equation*}
 \hat\gamma^{\theta*}_{h_t}(s')=\begin{cases}
 \frac{z^{\theta*}_{h_t}(s')}{\underset{\hat{s} \in \Aactsset(s_t)}{\sum}z^{\theta*}_{h_t}(\hat{s})}, \quad \text{if}\quad \underset{\hat{s} \in \Aactsset(s_t)}{\sum }z^{\theta*}_{h_t}(\hat{s})>0,\\
 \frac{1}{|\Aactsset(s_t)|},\quad\text{otherwise}.
 \end{cases} 
\end{equation*}

Defender's optimal strategy $\hat \pi^{*}$ is given by $\hat \pi^{*}_{h_t}$ in solution of the following dual LP:
\begin{align}\label{eq: Def. LP}
&\qquad \qquad \qquad \qquad\max_{\substack{\hat\Defs_{h_{t}},(q^{\theta}_{h_{t}})_{\theta \in \Theta},\\U,\; \forall h_t \in \restrictedH}} U\\ 
&\qquad \qquad \text{s.t. } \; \forall \theta \in \Theta, \; h_t \sqsubseteq h_{t+1}, \; h_t,h_{t+1} \in \restrictedH:\notag \\
&\qquad \qquad \mathbf{1}^{\top}\hat \pi_{h_t}=1,\;\hat \pi_{h_t}\geq\mathbf{0},\; \text{and} \; q^{\theta}_{h_0}\geq U\notag,\\
&L^{\theta\top}_{h_t} \hat\pi_{h_t}+\left[q^{\theta}_{h_{t+1}} \right]_{s_{t+1}\in \Aactsset(s_t)}\geq q^{\theta}_{h_t}\mathbf{1}, \; \forall h_t \in {\restrictedH \setminus \mathcal{H}^{\theta}_f}, \notag \\
&\qquad \qquad \qquad L^{\theta \top}_{h_t}\pi_{h_t}\geq q^{\theta}_{h_{t}}\mathbf{1},\quad \forall h_t \in {\mathcal{H}^{\theta}_f}.\notag
\end{align}
\noindent
Note that it is unnecessary to solve the LP~\eqref{eq: Def. LP} to find the solution for the Defender's strategy $\hat{\pi}^{*}$, one can extract the Defender's strategy from the dual variables for the optimality constraints of the primal LP~\eqref{eq. Attacker's LP}.

In the next subsection we lay out an algorithm that utilizes the LP in~\eqref{eq. Attacker's LP} and iteratively expands the restricted game.

\subsection{Default strategies.}\label{subsection unrestricted game}
\noindent
Default strategies specify the actions \emph{outside} the restricted game in order to specify the continuation values $\mathcal{U}^{\theta}_{h_{t+1}}$ at the leaf nodes, i.e. nodes corresponding to histories $h_t \in \overline{\mathcal{H}}_f^\theta$, (see~\eqref{eq. stage cost matrix}).
There are two key considerations for our selection of the default strategies. 
Firstly, we ensure that $\mathcal{U}^{\theta}_{h_t}$ are optimistic estimates of the true values from the Attacker's perspective, which is necessary for the optimality result presented in Proposition~1. To this end, we select the Attacker's default strategy to be a best response to the Defender's default strategy. 
Secondly, we select the Defender's strategy so that the aforementioned best response can be computed easily.

Defining the default strategies explicitly might take a large amount of memory, but in practice they can be defined using some general rule.
We set the Defender's default strategy, denoted as $\tilde\pi\in \Pi$, as allocation to the goal closest to the Attacker:
\begin{equation}
 \tilde{\pi}_{h_t}(\theta_\text{close})=1, \text{ where}\; \theta_\text{close}=\arg\underset{{\psi \in \Theta}}{\min}\; {d(s_t,\psi)}.
\end{equation}
In case of a tie, the Defender selects the goal with the smallest index.

The Attacker's default strategy is found as a deterministic best response to $\tilde{\pi}$. 
Since we selected $\tilde{\pi}$ to depend only on the Attacker's \emph{current} location, the Attacker's best response can be formulated as a shortest path problem on an MDP. 
The optimal strategy $\tilde{\gamma}^{\theta}_{h_t}$ and the value $V^{\theta*}(s), \; \forall s \in \mathcal{V}$ of this shortest path problem can be found via standard value iteration and the Bellman equation~\cite{filar2012competitive}.\footnote{To ensure that the Attacker reaches the goal in finite time, an arbitrarily small number can be added to the stage cost.}
This optimal value provides the continuation cost in the restricted game, $\mathcal{U}^{\theta}_{h_t}=V^{\theta*}(s_t)$.

\subsection{Expansion of the restricted game.}
\noindent
Algorithm~\ref{Algorithm1} summarizes the process to expand the restricted game based on the XDO algorithm.
Starting from the initial $\restrictedH$, we compute the Attacker’s strategy $\hat{\gamma}^{*}$ for the restricted game by solving LP~\eqref{eq. Attacker's LP}.
For each Attacker type $\theta$ played with positive probability, i.e., $\hat{\gamma}^{*}_{\text{type}}(\theta|s_0)>0$, we identify the reachable frontier histories
\begin{equation} \label{eq. histories in restrited game H,theta, gamma,f}
{\overline{\mathcal{H}}}^{\theta,\hat\gamma^*}_f
\coloneqq
\{h_t \in {\overline{\mathcal{H}}}^{\theta}_f
\mid
\mathbb{P}^{\hat\gamma^*}(h_t|\theta)\,\hat{\gamma}^{*}_{\text{type}}(\theta|s_0)>0\}.
\end{equation}
Each history $h_t \in \bigcup_{\theta \in \Theta}{\overline{\mathcal{H}}}^{\theta,\hat\gamma^*}_f$ is expanded by one time step. 
The resulting histories are added to $\restrictedH$, and the restricted game is solved again with the updated stage-cost matrices $L^{\theta}_{h_t}$. This process iteratively enlarges the restricted game and terminates when the optimal Attacker strategy no-longer leads to a non-terminal history, i.e.,
$\bigcup_{\theta \in \Theta}{\overline{\mathcal{H}}}^{\theta,\hat\gamma^*}_f=\{\emptyset\}.$
\begin{algorithm} 
	\caption{Solve unrestricted Game}\label{Algorithm1} 
	\begin{algorithmic}[1]
\Require initial $\restrictedH$,
\Statex \quad \; \quad $\hat\gamma^{*}$ for initial $\restrictedH$ given by LP~\eqref{eq. Attacker's LP}
\State For all $\theta \in \Theta$ find ${\overline{\mathcal{H}}}^{\theta,\hat\gamma^*}_f$~using~\eqref{eq. histories in restrited game H,theta, gamma,f}
\While {$\bigcup_{\theta \in \Theta}{\overline{\mathcal{H}}}^{\theta,\hat\gamma^*}_f\neq \{ \emptyset \}$} \Comment{Non-terminal history exists}
 
		\For { $h_t\in \bigcup_{\theta \in \Theta} {\overline{\mathcal{H}}}^{\theta,\hat\gamma^*}_f$ }
 \For{$s_{t+1} \in \mathcal{N}(s_t)$} 
 \State $h_{t+1}=(h_{t},s_{t+1})$ \Comment{Create a new history}
 \State $ \restrictedH\leftarrow \restrictedH\cup \{h_{t+1} \}$ 
			\EndFor
 \EndFor
 \State For every type $\theta\in \Theta$ reassign $L^{\theta}_{h_t}$ for every $h_t\in \restrictedH$
			\State Solve for $\hat\gamma^{*}$ using LP~\eqref{eq. Attacker's LP}
 \State For all $\theta \in \Theta$ find new ${\overline{\mathcal{H}}}^{\theta,\hat\gamma^*}_f$~using~\eqref{eq. histories in restrited game H,theta, gamma,f}
			\EndWhile
 \State Solve for $\hat \pi^{*}$ using LP~\eqref{eq: Def. LP}
 \State \Return $\hat \pi^{*}$, $\hat \gamma^{*}$, $\restrictedH$
	\end{algorithmic} 
\end{algorithm}

Algorithm~\ref{Algorithm1} is conceptually similar to the $\text{A}^{*}$ search algorithm~\cite{hart1968formal}. The continuation cost $\mathcal{U}^{\theta}_{h_t}$ acts as an optimistic heuristic estimate of the remaining cost for the Attacker. It never overestimates the true equilibrium expected cost because $\mathcal{U}^{\theta}_{h_t}$ is computed as the Attacker’s best response to the Defender’s default strategy.
Therefore, histories that are not reached with positive probability under the current optimal strategy $\hat{\gamma}^*$ cannot reduce the Attacker’s equilibrium cost and can be safely ignored.

The equilibrium Attacker's strategy is given by $\hat{\gamma}^{*}$, obtained from Algorithm~\ref{Algorithm1}, i.e., $\gamma^{*}=\hat{\gamma}^{*}$. The Defender's equilibrium strategy is given by $\hat{\pi}^{*}$ from Algorithm~1 and the default strategy $\tilde \pi$:
\begin{equation}
 \pi^{*}_{h_t}(\theta)=\begin{cases}
 \hat \pi^{*}_{h_t}(\theta), \; \text{if} \; h_t\in \restrictedH,\\
 \tilde \pi_{h_t}(\theta), \; \text{otherwise}.
 \end{cases}
\end{equation}

\begin{prop}
 The output of Algorithm~1 $(\pi^{*},\gamma^{*})$ constitutes a Nash equilibrium of the original unrestricted game: i.e., it satisfies~\eqref{eq. Nash}.
 
\end{prop}
\begin{Proof}
The linear programs~\eqref{eq. Attacker's LP} and~\eqref{eq: Def. LP} produce strategies for both players that constitute a Nash equilibrium in the restricted game by construction. 

Moreover, for any Attacker strategy $\hat{\gamma}$ such that $\mathbb{P}^{\hat{\gamma}}(h_t)=0$ for all $h_t \in \overline{\mathcal{H}}_f^\theta$, the cost in the restricted game coincides with the cost in the full game for any Defender policy. Since the Attacker alone controls the evolution of the game history, the probability $\mathbb{P}^{\hat{\gamma}}(h_t)$ is independent of the Defender’s strategy. Consequently, the Defender cannot force the game to reach any history $h_t \in \overline{\mathcal{H}}_f^\theta$, and thus the Defender’s best response in the restricted game is also a best response in the full game.

It therefore remains to show that the Attacker has no incentive to use a strategy that generates histories $h_t \in \overline{\mathcal{H}}_f^\theta$. This property holds once Algorithm~\ref{Algorithm1} converges. The algorithm terminates when the \textbf{while} loop exits, which occurs when every selected Attacker type reaches its intended goal in the restricted game, i.e., when $\mathbb{P}^{\hat{\gamma}}(h_t)=0$ for all $h_t \in \overline{\mathcal{H}}_f^\theta$. 

Under this condition, any deviation by the Attacker that drives the game to a history $h_t \in \overline{\mathcal{H}}_f^\theta$ incurs a cost that is at least the continuation cost computed for that history. This continuation cost is evaluated assuming a best response by the Attacker to the Defender’s default strategy. Since the Defender indeed uses the default strategy for such histories under $\pi^{*}$, the Attacker cannot obtain a lower cost in the full game by deviating in this manner. Therefore, the Attacker has no incentive to generate histories in $\overline{\mathcal{H}}_f^\theta$, and $(\pi^{*},\gamma^{*})$ constitutes a Nash equilibrium of the full game.
\end{Proof}

Note that Algorithm~\ref{Algorithm1} is guaranteed to terminate in finite time if there exists an equilibrium strategy for Attacker such that all histories generated by it reach the goal in bounded time. Although there may be cases where Attacker's equilibrium strategy must generate histories which do not reach the goal for an arbitrarily long time, this has not occurred in 
any tested examples, such as those in Section~\ref{sec:numerical}, and we conjecture that this does not occur.




\subsection{Discussion}
\noindent
In equilibrium, the Attacker may assign zero probability to some types $\theta$, i.e., $\gamma^{*}_{\text{type}}(\theta|s_0)=0$, because the corresponding goal may not be worth pursuing.
Suppose the equilibrium Attacker selects multiple types $\theta,\psi \in \Theta$ with positive probability, i.e., 
$\gamma^{*}_{\text{type}}(\theta|s_0)>0$ and $\gamma^{*}_{\text{type}}(\psi|s_0)>0$. 
Then the equilibrium expected costs for these types must be equal. 
Otherwise, if 
$J^\theta(\pi^*,\gamma^{\theta*};s_0,\mathbf{r})
<
J^\psi(\pi^*,\gamma^{\psi*};s_0,\mathbf{r})$,
the Attacker could strictly decrease its expected cost by assigning higher probability to type $\theta$ and zero probability to type $\psi$, contradicting the optimality of $\gamma^*$. 

A similar argument applies to trajectories. 
Given $\gamma^{*}_{\text{type}}(\theta|s_0)>0$, all terminal histories $h_t \in \mathcal{H}^{\theta*}_f$ realized by $\gamma^{\theta*}$ must yield the same equilibrium expected cost. 
Otherwise, the Attacker could reduce its cost by shifting probability to a trajectory with a lower cost. 
Thus, the equilibrium Defender makes the Attacker indifferent between its equilibrium type(s) and equilibrium trajectories.

\begin{rem}[On-path indifference]
Let $
\hat\Theta
\coloneqq
\{\theta \in \Theta \mid \gamma^*_{\textnormal{type}}(\theta|s_0)>0\}$.
Then:
\begin{enumerate}[label=(\roman*)]
\item {Indifference between equilibrium types:} 
For any $\theta,\psi \in \hat\Theta$,
\[
J^\theta(\pi^*,\gamma^{\theta*};s_0,\mathbf{r})
=
J^\psi(\pi^*,\gamma^{\psi*};s_0,\mathbf{r}).
\]

\item {Indifference between equilibrium trajectories:} 
For any $\theta \in \hat\Theta$, all terminal histories 
$h_t \in \mathcal{H}^{\theta*}_f$ that occur with positive probability under $\gamma^{\theta*}$ yield the same equilibrium expected cost.
\end{enumerate}
\end{rem}

In the next section we propose two metrics to quantify deceptive behavior of the Attacker.

\section{Deception metrics}

\noindent
To evaluate deceptive behavior in the incomplete information game, we first establish the benchmark value of the \emph{complete information game}, when the Defender knows the Attacker's selected goal $\theta$ at the start of the game. This benchmark allows us to define two metrics: a) Risk of Deception (RoD) -- the expected penalty incurred by the Attacker when a deceptive strategy is used against a fully informed Defender; and b) Value of Information (VoI)~\cite{HOWARDvoi} -- the relative advantage the Attacker gains from private information compared to the complete information case.
Together, these metrics capture the trade-off between risk and benefit in deceptive play. 

\paragraph*{Complete Information Game}

Suppose the Defender knows the Attacker’s type $\theta$ at the start of the game. In this case, the Defender’s optimal strategy is to allocate to $\theta$ every time step, while the Attacker’s best response is to follow the shortest path to its goal. The cost to the Attacker $\theta$ is therefore the sum of the initial allocation and the distance to $\theta$ (length of the shortest path to $\theta$). 
If the Attacker anticipates that its type will be revealed after selection, it will always choose the type minimizing this cost.
\begin{rem}[Complete Information Game Value]
The value of the complete information game for type $\theta$ is
\begin{equation}
 \bar U^{\theta*}(s_{0},r^\theta) = d(s_{0}, \theta) + r^{\theta},
\end{equation}
and the overall value is
\begin{equation}
 \bar U^{*}(s_{0},\mathbf{r}) = \min_{\theta \in \Theta} \bar U^{\theta*}(s_0,r^\theta).
\end{equation}
\end{rem}

This benchmark provides a natural reference point for defining both RoD and VoI.

\paragraph*{Risk of Deception (RoD)}

In complete information game, the Attacker's deceptive strategy that deviates from a shortest path will be suboptimal. Consequently, when the Attacker is unsure whether the Defender does or does not know its private type, the Attacker faces a dilemma: (i) attempt at deceiving the Defender; or (ii) remain on the shortest path.
If the Defender knows $\theta$, then the cost of playing $\gamma^{\theta}$ is the expected path length, i.e.:
\begin{equation*}
 \hat d^{\gamma^{\theta}}(s_0,r^{\theta}) \coloneqq \mathbb{E}^{\gamma^{\theta}}\!\left[\sum_{t=0}^{T^{\theta}-1} w(s_t,s_{t+1}) + r^{\theta}\right].
\end{equation*}

\begin{definition}
The risk of deception for Attacker $\theta $ is the excess cost of playing $\gamma^{\theta *}$ instead of staying on a shortest path when the Defender knows $\theta$:
\begin{equation*}
 \textnormal{RoD}^{\theta}(s_0,\mathbf{r}) 
 = \hat d^{\gamma^{\theta*}}(s_0,r^{\theta}) - \bar U^{\theta*}(s_0,\mathbf{r}).
\end{equation*}
The expected risk of the overall strategy $\gamma^*$ is
\begin{equation*}
 \textnormal{RoD}(s_0,\mathbf{r}) 
 = \sum_{\theta \in \Theta} \gamma_{\textnormal{type}}^{*}(\theta|s_0)\,\textnormal{RoD}^{\theta}(s_0,\mathbf{r}).
\end{equation*}
\end{definition}

If $\text{RoD}^{\theta}(s_0,\mathbf{r})=0$, it implies that Attacker type $\theta$ always remains on a shortest path. 
Likewise, $\text{RoD}(s_0,\mathbf{r})=0$ implies that all types selected with non-zero probability stay on shortest paths. 
Positive RoD values quantify the expected penalty that the Attacker faces when attempting deception against an informed Defender.

\paragraph*{Value of Information (VoI)}

VoI~\cite{HOWARDvoi} quantifies the (normalized) advantage that the Attacker gains from having a private information, relative to the complete information benchmark.

\begin{definition}
For Attacker $\theta$, the value of information is
\begin{equation*}
 \textnormal{VoI}^{\theta}( s_0,\mathbf{r}) 
 =  \frac{\bar U^{\theta*}(s_0,\mathbf{r}) - J^{\theta}(\pi^{*},\gamma^{\theta*};s_0,\mathbf{r})}{\bar U^{\theta*}(s_0,\mathbf{r})}.
\end{equation*}
The overall value of information is
\begin{equation*}
 \textnormal{VoI}( s_0,\mathbf{r}) 
 =   \frac{\bar U^{*}(s_0,\mathbf{r}) - J(\pi^{*},\gamma^*;s_0,\mathbf{r})}{\bar U^{*}(s_0,\mathbf{r})} .
\end{equation*}

\end{definition}

A positive value of information $\textnormal{VoI}(s_0,\mathbf{r}) > 0$ indicates that private information will provide a positive benefit to the Attacker. 
Note that $\bar U^{\theta*}(s_0,\mathbf{r})=0$ occurs only in the degenerate case where the Attacker starts at goal $\theta$ with zero initial allocation, i.e., $r^\theta=0$, and the game immediately ends.
Thus we assume that the denominator is nonzero.  Otherwise, we define $\text{VoI}(s_0,\mathbf{r})=0$. 

The two measures are complementary: RoD highlights the vulnerability of deceptive strategies when they are revealed, while VoI highlights the potential benefit of deception when the Attacker's type remains private.

\section{Numerical Illustration}\label{sec:numerical}
\begin{figure}[h]
 \centering
 \includegraphics[width = 0.47\textwidth]{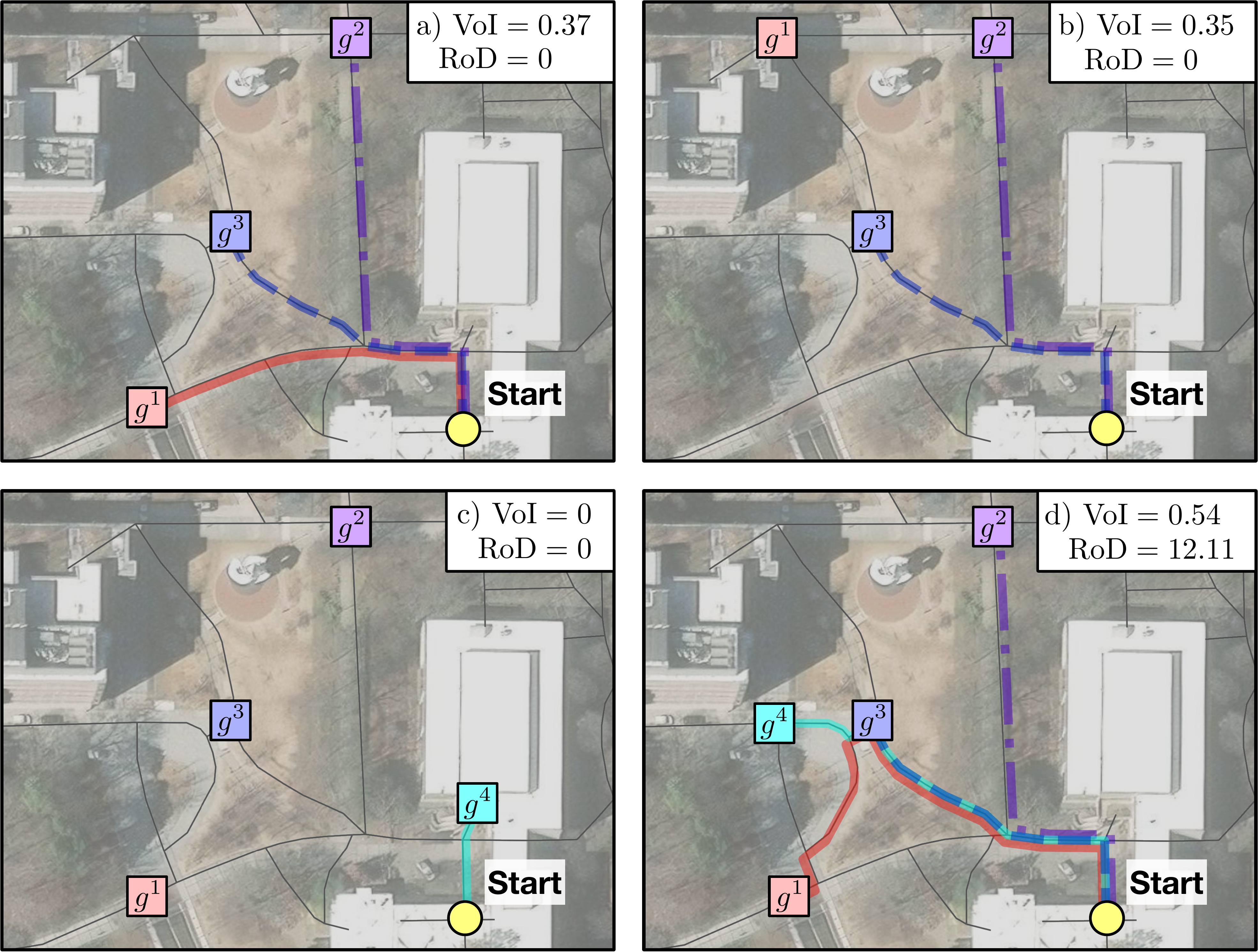}
 \caption{Equilibrium strategies for different initial conditions. Equilibrium cost is: a) 61.54, b) 63.15, c) 48.52, d) 45.30.}\label{fig: figure_2}
\end{figure}
\noindent We illustrate the equilibrium solution in two settings: a realistic road network and a grid world. 
We highlight how optimal deception emerges and how it depends on environment structure and initial conditions. Across both settings we contrast outcomes under asymmetric and complete information by showing the value of the game, VoI, and RoD. 

\subsection{Road network.}
\noindent
Figure~\ref{fig: figure_2} illustrates scenarios on a road-network graph extracted from real map data and presents \(\text{VoI}\coloneqq \text{VoI}( s_{0}, \mathbf{r})\) and \(\text{RoD}\coloneqq \text{RoD}(s_{0}, \mathbf{r})\). Across all cases we set the initial allocations to zero, i.e., $r^{\theta}=0$, $\forall \theta\in\Theta$.

In Fig.~\ref{fig: figure_2}{a}, the \Attacker{} mixes uniformly over all goals and, conditional on the chosen goal, follows a shortest path toward it. 
In Fig.~\ref{fig: figure_2}{b} goal \(g^{1}\) is relocated. The equilibrium \Attacker{} never pursues \(g^{1}\) and randomizes uniformly over \(g^{2}\) and \(g^{3}\). Consequently, \(\text{VoI}\) decreases relative to Fig.~\ref{fig: figure_2}{a} because the path to \(g^{1}\) no longer aids concealment. Note that $\text{RoD}=0$ in both Fig.~\ref{fig: figure_2}{a} and~\ref{fig: figure_2}{b}, which means that the paths taken by the equilibrium Attacker are the shortest ones, yet $\text{VoI}>0$. This highlights the benefit the equilibrium Attacker receives from the Defender's informational disadvantage.

Figures~\ref{fig: figure_2}{c} and~\ref{fig: figure_2}{d} add a new goal \(g^{4}\). On Fig.~\ref{fig: figure_2}{c} the \Attacker{} selects only \(g^{4}\) and follows the shortest path. Here \(\text{VoI}=0\), yet the \Attacker’s total cost is lower than in Fig.~\ref{fig: figure_2}{a}, illustrating that \(\text{VoI}\) does not directly track the equilibrium payoff.
In Fig.~\ref{fig: figure_2}{d}, the placement of \(g^{4}\) induces uniform mixing over all four goals, with routes to \(g^{1}\) and \(g^{4}\) transiting through \(g^{3}\). Here the \Attacker{} chooses a non-shortest path to \(g^{1}\), yielding \(\text{RoD}>0\), while \(\text{VoI}\) has the highest value among the four cases.
\begin{figure}[h!]
 \centering
 \includegraphics[width = 0.45\textwidth]{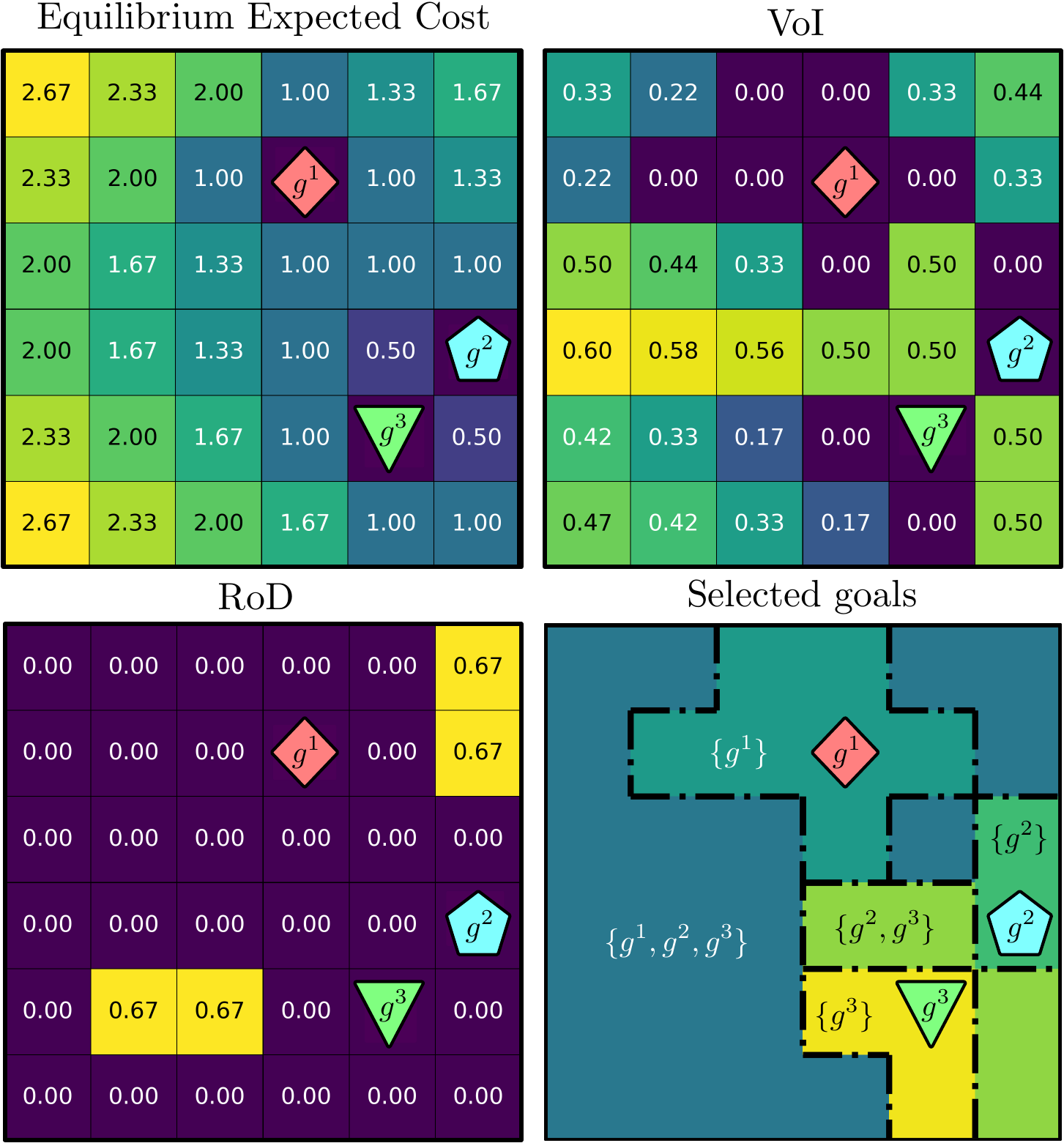}
 \caption{Equilibrium values for expected cost (top left), $\text{VoI}$ (top right), $\text{RoD}$ (bottom left), and set of selected goals, i.e. $\gamma^{*}_{\text{type}}(\theta|s_0)>0$ (bottom right). }\label{fig: figure: 3}
\end{figure}
\subsection{Grid world.}
\noindent
Figure~\ref{fig: figure: 3} presents a grid-world scenario with three goals. All values depict equilibrium performance under $r^{\theta}=0$, $\forall\theta\in\Theta$. The bottom-right subfigure shows that when the \Attacker{} starts close to a particular goal, it often commits to that goal alone. The game then becomes equivalent to a complete-information case, which is reflected by VoI being zero, as shown in the top-right subfigure. 
In contrast, the dark-green region (labeled as $\{g^1,g^2,g^3\}$) away from all three goals lead to mixing between all three goals.
Concealing the intent provides a tangible advantage captured by positive $\mathrm{VoI}$, and the \Attacker{} often accepts non-shortest path detours, as evidenced by positive $\mathrm{RoD}$ at certain initial positions.
\balance
\section{Conclusion}
\noindent
We formulated a new variant of the deceptive path planning~(DPP) game with goal selection, and introduced a solution method that integrates linear programming with the Double-Oracle algorithm. 
We propose a particular choice of default strategies that allow us to solve DPP game which has indefinite horizon, and it also ensures our solution is an exact Nash equilibrium, rather than an $\varepsilon$-Nash.
We present two complementary metrics to quantify the benefits and costs of deception: Value of Information (VoI) and Risk of Deception (RoD). Together, these measures delineate conditions where deception is both effective and risk-averse.
The illustrative examples reveal how deceptive behavior can improve performance. 
Future work will address scenarios with partial observability, stochastic dynamics and games in continuous space.


\bibliographystyle{IEEEtran}
\bibliography{ref}

\end{document}